\documentclass[prb,twocolumn,showpacs,aps]{revtex4}

\begin{document}

\title{Fingerprinting the Electronic
Wavefunctions of Ultra-Small Conductors}

\author{Gustavo A. Narvaez} 
\altaffiliation[Present address:~]{Department of Physics, The Ohio State University, Columbus, Ohio 43210}
\author{George Kirczenow}


\affiliation{Department of Physics, Simon Fraser University, Burnaby, British
Columbia, Canada V5A 1S6}

\date{\today}
%

\begin{abstract}
By extending the Orthodox theory of Coulomb blockade to include
the nano-scale effects of environmental electric fields, we show that
tunneling spectra of metallic
nanoislands contain information not only on the energies of the electronic
levels but also on their {\em wavefunctions} near the surface of the island.
This fundamental additional
information is predicted to take the form of new
observable phenomena beyond the scope of standard
Orthodox theory: the tunneling resonances are renormalized and show
surprisingly strong Gaussian fluctuations, level bending and avoided crossings.
\end{abstract}

\maketitle

\section{Introduction}
%
%
The Orthodox theory of Coulomb blockade\cite{CB} is a cornerstone of
the present understanding of charge transport in single-electron
devices\cite{set,set1}. In this theory the device
(a small metal, semiconductor or fullerene island and electrodes) is modeled
using an equivalent electric circuit with macroscopic
components (capacitors and resistors). Electronic interactions that
control the passage of charge through the island are accounted
for classically by an electrostatic {\em
charging} energy
$U_c$. The quantum nature of the island is incorporated through
the discreteness of its energy levels. 
The statistical properties of the electron wavefunctions and energy levels are
usually described within the frame of random matrix
theory\cite{alhassid_RMP_2000,RMT_reviews,brody_RMP_1981} or by
exact diagonalization of a microscopic Hamiltonian\cite{narvaez_PRB_2002}.
The
average energy level separation
$\delta$ and $U_c$ set the energy scales of the transport problem. The
Orthodox theory works impressively well for mesoscopic islands.
Nevertheless, recently, a new generation of nanoscopic metallic islands ({\em
nanoislands}) has yielded surprising tunneling spectroscopy
results\cite{nanometals,petta_PRL_2001,ralph_experiments,ralph_PRL_1997}
which have led theoreticians to extend the Orthodox theory in order to
successfully account
\cite{agam_PRL_1997,g-factor_models,theories,vondelft_PhysRep_2001} for
some of the novel observations.

%
%
In this paper, the Orthodox
theory is extended further in order to explore the
microscopic effects of environmental electric fields on
the tunneling spectroscopy of these nanoislands. These effects are due to
penetration of the fields into the island, therefore they become  more
significant as the size of the island is reduced. To our knowledge
they have not been considered previously in the experimental
or theoretical literature. Our paradigm for
studying them will be the  transistor introduced by Ralph,
Black and Tinkham\cite{ralph_PRL_1997}, although environmental fields
should affect all nanoscopic single-electron transistors.
%
%
We show that environmental fields result in an {\em electrostatic
renormalization} of the electronic states of the nanoisland that is
beyond the scope of the standard Orthodox
theory.
This renormalization exhibits nearly-Gaussian fluctuations
due to the stochastic nature of the electron
wavefunctions. Furthermore,
we predict the occurrence of avoided crossings and level
bending due to the coupling between electronic levels
induced by the fields. We show that these
phenomena are reflected directly in the behavior of
tunneling resonances; this establishes a platform for
probing the environmental coupling experimentally.
The fluctuations in the slopes of
the tunneling resonances vs. gate voltage that we predict are
surprisingly strong, comparable in size to the unrenormalized slope
predicted  by Orthodox theory. We demonstrate that by observing these
fluctuations it is possible to assess the amplitudes of the
electronic wavefunctions at the surface of an ultra-small conductor.
Moreover, the statistics of the wavefunctions at the surface of the
nanoparticle can be separated from those in the bulk and measured. No
way to measure these important properties
of nanoparticle electron wave functions has been known until now.
We conclude that a
novel fingerprint of the electron wavefunctions at the surfaces
of nanoscopic conductors can be found in tunneling
spectroscopy experiments.

%
%
\section{\label{environmental}Environmental electric fields}
We now introduce the environmental electric fields.
The model geometry\cite{ralph_PRL_1997} is shown in Fig.
\ref{Fig_1}(a): Three macroscopic electrodes labeled
$i=L$ and $R$ (the source and drain contacts) and $G$ (the gate), are
separated from the central nanoisland
      ($P$) by thin insulating layers of thickness $d^i$,
and are connected to voltage sources
$V_L$,
$V_R$, and
$V_G$ which drive the island to a potential
$
V_P=(C_{\Sigma})^{-1}\left[Q_P-(C_L - C_R)V/2+C_G V_G\right]
$
in the Orthodox theory. $Q_P$ is the charge  on the nanoisland.
$C_L$, $C_R$, and $C_G$ are capacitances between electrode $i$ and the
nanoisland, and $C_{\Sigma}=C_L+C_R+C_G$.
$V=V_R-V_L$ is the bias voltage with
$-V_L=V_R=V/2$.
The  difference between $V_i$ and $V_P$ sets an electric field at the
nanoisland surface ($\Omega_i$)
      that faces electrode {\em i}.
The electric field has a
finite penetration depth ($d_s$) into the conductor (in the standard Orthodox
model $d_s = 0$) that causes the
electrostatic potential within the nanoisland to deviate from $V_P$.
This deviation ($\Delta V^i_z=V^i_z-V_P$) is sketched in Figure
\ref{Fig_1}(b). Assuming
that the electric field within the metal nanoisland is screened exponentially\cite{ashcroft_book},
$\Delta V^i_z$ at a distance $z$ along the direction normal to $\Omega_i$ can
be estimated as
$
\Delta V^i_z={\alpha}_i (\lambda d^i+2)^{-1}(V_i-V_P) e^{-\lambda z}.\cite{note_10}
$
${\alpha}_i=\epsilon(\lambda d^i+2)/(\lambda d^i +2\epsilon)$
is a polarization enhancement factor, where
$\epsilon$ is the dielectric constant of the insulating
material, and $\lambda=1/d_s$ ($\simeq 1$\AA$^{-1}$ for 
Al\cite{newns_PR_1970}).
Due to the effective screening (small $d_s$) in metals,
only atoms near the surface
are significantly perturbed by the electrostatic potential energy
${\cal{U}}=-e\sum_{i}\Delta V^i_z$.\cite{note_14} However, in the {\em nanoscopic} regime
the substantial surface-to-volume ratio ($N_{\Omega}/N_{{\cal V}}\sim 1$) of
the nanoisland warrants the inclusion of
${\cal{U}}$ in the calculation of its electronic structure.
${\cal{U}}$ causes what we call the {\em microscopic electrostatic
renormalization} of the energy levels, that is not included
in the standard Orthodox theory.
This renormalization is {\em in addition} to
the {\em bulk} or {\em macroscopic} one due to
$-eV_P$ that is included in the standard Orthodox theory. 

%
%
\section{Results and discussion}
We begin by exploring the effects of ${\cal{U}}$ analytically
within perturbation theory. We then present results of our {\em
non-perturbative} computer simulations that apply to specific real systems:
Al nanoislands with oxide tunnel barriers.
%
%
%
%
\subsection{\label{perturbation}Perturbation theory in ${\cal U}$}
We define the {\em microscopic} contribution to the
energy ${\cal E}^q_a(V,V_G)$ of level
$|\psi_a\rangle$ to be
$\varepsilon^{q}_a(V,V_G)$, where $q=Q_P/e$ labels the charge state of
the nanoisland.
Let $|\psi^0_a\rangle$  be an electronic eigenstate for $V_G=0$.
For $V_G \ne 0$ these states become {\em
coupled}:
${\cal U}_{ab}=\langle\psi^0_a|{\cal U}|\psi^0_b\rangle\neq 0$.
Now suppose the coupling
strength $\gamma_{ab}={\cal U}_{ab}/[{\cal E}^0_a(0,0)-{\cal
E}^0_b(0,0)]$ is small for all $|\psi^0_b\rangle$. Then, to first
order in ${\cal U}$, (setting for simplicity $V=q=0$)
$\varepsilon^0_a(0,V_G)$ is {\em linear} in $V_G$ with slope (see Appendix \ref{derivation_0})
%
%
%
\begin{widetext}
\begin{equation}
{
\begin{array}{rl}
S^{m}_{a} = \delta \varepsilon^{0}_a(0,V_G)/(e\delta V_G)  &
=
\,-(C_G/C_{\Sigma})
\sum_{j}|\psi^0_a(j)|^2 \left[-\alpha_{L} (\lambda d^L+2)^{-1}
e^{-\lambda z^{L}_j}
    - \alpha_{R} (\lambda d^R + 2)^{-1} e^{-\lambda z^R_j}
\right. \\
    & \left. +
(C_{\Sigma}/C_G-1)\alpha_{G} (\lambda d^G+2)^{-1} e^{-\lambda z^G_j}  \right]
\end{array}
},
\label{Eq_slopes}
\end{equation}
%
\end{widetext}
%
%
%
%
\noindent where $|\psi^0_a(j)|^2$ is
the amplitude of $|\psi^0_{a}\rangle$ at atomic site $j$ of the nanoisland.
$z^{i}_{j}$ is the distance of site $j$ from the surface $\Omega_i$.
Because $d_s$ is smaller than the size of the nanoisland
the leading terms in the summations correspond to atomic sites on
the surfaces $\Omega_i$. The
surface of the nanoisland presents atomic scale disorder.
Consequently, in the absence of magnetic fields, the electronic structure
is described by the Gaussian orthogonal ensemble of random matrix theory
\cite{RMT_reviews,narvaez_PRB_2002}. Therefore,
$|\psi^0_a(j)|^2$ is a random variable that obeys a
$\chi^2$ distribution \cite{brody_RMP_1981}. Thus the slope
$S^{m}_{a}$ {\em fluctuates} from state to state\cite{note_2} reflecting
the microscopic details of the electron wavefunctions at the surfaces of the
nanoisland. Due to the form of Eq. (\ref{Eq_slopes}) the fluctuations are
nearly-Gaussian according to the central limit theorem. These observations
remain valid for arbitrary
$V$ and $q$. We conclude from this
analysis that the microscopic effects of the environmental
electric fields lead to a fluctuating renormalization ({\em
microscopic electrostatic renormalization}) of the electronic levels
of a conducting nanoisland. It should be noted that in the standard Orthodox
model ($d_s=0$) $S^m_a$ is zero.
%
%
%

Consider now the implications of this microscopic effect for the
tunneling spectroscopy of the nano-transistors.  Suppose  $Q_P=0$
at $V=0$ and fixed $V_G$, and then, after increasing
the bias, an
electron  tunnels in from  electrode $L$ ($L\rightarrow
P$) so that $Q_P=-e$. If $|\psi_a\rangle$ is being populated by the
tunneling electron the  transition is favorable only if the final state
energy of the electron is lower than
${\cal E}_F$, the Fermi energy of electrode $L$:
$
{\cal E}_F \geq {\cal E}^{-}_a(V,V_G)=\varepsilon^{-}_a(V,V_G)+ U_c -
[(C_R + C_G/2)/C_{\Sigma}]eV
-(C_G/C_{\Sigma})e V_G$, where $U_c=e^2/2 C_{\Sigma}$ is the
charging energy\cite{CB}.
(Note that the {\em macroscopic}  renormalization due to
$Q_P$ (through $U_c$), $V$ and $V_G$
constitutes the Orthodox model.)
A similar restriction applies to tunneling transitions from the  neutral
nanoisland to electrode
$R$.
These conditions for tunneling set the threshold biases ($V^{th}$),
the minimum values of $V$ that satisfy them
for a given
$|\psi_a\rangle$.
As $V$ is swept at fixed
$V_G$, each time a new $V^{th}$ is
reached a peak (a threshold
resonance ({\em TR})) appears in
the differential conductance
$dI/dV$\cite{averin_JETP_1990}.
We now
analyze the effect of  $V_G$ on the {\em TR},
focusing on  $L\rightarrow P$ transitions.
In the perturbative regime we find (see Appendix \ref{derivation})

%
%
\begin{widetext}
\begin{equation}
\frac{\delta V^{th}_{L\rightarrow P}(V_G)}{ \delta V_G} =
-\frac{C_G}{C_R+C_G/2}
\left\{
1
+S^{m}_{a} (-C_G/C_{\Sigma})^{-1}
-J^{m}_{a} [-(C_R+C_G/2)/C_{\Sigma}]^{-1}
+\dots
\right\},
\label{slopes}
\end{equation}
%
%
\end{widetext}
%
%
%
%
\noindent where
$J^m_a=\delta \varepsilon^{-}_a(V,V_G)/(e \delta
V)$\cite{note-3}. Equation (\ref{slopes}) shows that the positions
of the resonances in the
$dI/dV$ spectrum shift linearly with $V_G$. The first term on
the right in Eq. (\ref{slopes}) is the ({\em macroscopic})
slope predicted by standard Orthodox
theory\cite{vondelft_PhysRep_2001};
%
%
the other
terms are due to  microscopic effects.
This analysis shows that the expected {\em
macroscopic} slope of the {\em tunneling resonances} is {\em microscopically}
renormalized. Furthermore, the slope of each resonance in the spectrum is
different as a consequence of the {\em fluctuations}. Therefore it
should be possible to study the coupling of the nanoisland's electronic
levels to the environment by tunneling spectroscopy, and thus
obtain experimental information on the electron wavefunction.

%
%
%
\subsection{Non-perturbative computer simulations}
So far our exposition, although perturbative, has been general. To understand
the effects of the electric fields on the electronic structure and tunneling
spectroscopy {\em beyond perturbation theory}, we now consider Al nanoislands
coated with Al-oxide within a tight-binding model
\cite{narvaez_PRB_2002}  with the on-site orbital energies of the
isolated neutral Al/Al-oxide nanoisland adapted to include ${\cal{U}}$.
The presence of the oxide coating introduce surface disorder in these nanoislands, as explained in 
Ref. \onlinecite{narvaez_PRB_2002} (see also Ref. \onlinecite{note_12}).
We consider disc-shaped nanoislands of
volume  ${\cal V}\simeq 17nm^3$, and
$d^{R}=10$\AA, $d^L=15$\AA and
$d^G=4\,d^L.$\cite{note-2} The latter choice
assures that tunneling from the gate is suppressed.
The parameters for two such systems are 
shown in Table \ref{tab:Table_1}. The Fermi energies of the neutral 
isolated nanoislands
($E^0_F$) differ due to their differing structural disorder and
geometries. The latter are also responsible for the differing capacitances.
The {\em charging} voltage
$V_U=e/C_{\Sigma}$ is the upwards (downwards) shift of
$V_P$ as one electron is removed (added) from (to) the nanoisland; the {\em
degeneracy} gate voltage $V^{\pm}_G=\pm e/(2 C_G)$ is,
within the standard Orthodox theory\cite{CB}, the value of $V_G$ at
which an electron is added ($+$) to or removed ($-$) from the nanoisland at
$V=0$.

Figure
\ref{Fig_2} shows results of the exact diagonalization
of the tight binding Hamiltonian for $\varepsilon^{q}_a(V,V_G)$ for a few
energy levels of transistors
$A$ and
$B$ under different
      bias and gate conditions, at $q=0$.
$V_G$ ranges from approximately $V^{-}_G$ to  $V^{+}_G$.
$V$ lies within the first step of the Coulomb staircase (at
$V_G=0$) of the nanoisland.
Let us first focus
on the effects of $V_G$ at $V=0$.
The striking features are: i) {\em fluctuating}
renormalization of the slopes of the energy levels due to the applied
electric field, ii) linear dependence of some energy levels on $V_G$,
and iii) level bending and avoided crossings with increasing $|V_G|$.
These features are qualitatively similar regardless the values of $q$
or $V$ (not shown).
We now proceed to discuss them. i) and ii) can be
understood within the perturbative argument presented above. Regarding iii)
it should be noted that as
$|V_G|$ increases significant enhancement of
$\gamma_{ab}$ can occur due to increased coupling between
levels and level proximity induced by $V_G$. In this case, the
microscopic contribution to the energy levels should be calculated to higher
order in
${\cal U}$. Level bending signals the response of the energy levels to this
change in the electronic coupling. As
$|V_G|$ increases some energy levels become nearly degenerate. Since these
levels are in general mixed by ${\cal{U}}$ this results in avoided
crossings in the spectrum. Thus level bending and avoided crossings
as a function of gate bias should be general features of the spectra of these
nano-transistors: Electric fields should effectively couple quasi-particle
levels (which may even involve electronic excitations) that belong to the
same Hilbert space {\em and} whose wavefunctions are such that the strength
parameter
$\gamma_{ab}$ is significant.
The microscopic effect of the source-drain voltage $V$ on the electronic structure is
also shown in Fig. \ref{Fig_2}  (for $q=V_G=0$); there are again
fluctuations and avoided crossings. These features are similar for finite 
values of $q$ and $V_G$ (not shown). The
renormalization as a function of $V$, however, is smaller than for $V_G$ due to the
electrode geometry.

In conclusion, we have shown that the electronic levels are sensitive to the
environmental electric fields and different
regimes can be identified depending on the
strength of the electric coupling between levels. The novel effects are
predicted to occur in realistic Al/oxide transistors at the meV
scale (see Fig. \ref{Fig_2}) that is readily accessible
experimentally.


We now  discuss the fluctuations of the slopes
using the calculated $\varepsilon^0_a(V,V_G)$. Table \ref{tab:Table_1} shows 
the average value (${\cal S}$) of the slopes
$S^m_a$, taken over
$n=101$ consecutive levels distributed symmetrically with respect to
$E^0_F$, and the Orthodox slope
$-C_G/C_{\Sigma}$  for comparison. ${\cal S}$ for
    transistor $A$ is greater than for $B$ by a factor of
$1.34$. This is consistent with  $\Omega_G$ being $1.38$ times bigger
in $A$ than in $B$ (see Table
\ref{tab:Table_1}) and, consequently, more atomic sites being perturbed by the
gate electric field.  The standard deviation ($\sigma$) of
      ${\cal S}^{m}_a$ is roughly {\em
equal} ($\simeq 2.3\times10^{-3}$) for the two transistors.
Figure \ref{Fig_3} shows the histograms for $S^{m}_a$ in units of
$-C_G/C_{\Sigma}$. The {\em microscopic} slopes of
the energy levels are clearly significant fractions of the {\em macroscopic}
slope arising from the Orthodox model. Indeed, for
$B$ the {\em microscopic} and the {\em macroscopic} effects are roughly
equal. Thus the
microscopic effects of environmental fields on the electronic structure and
tunneling resonances that we have derived here can be a {\em very} important
correction to macroscopic effects predicted by standard Orthodox
theory. Both distributions in Fig.
\ref{Fig_3} are nearly Gaussian (the smooth lines in Fig.
\ref{Fig_3} are
$n\Delta s (\sigma\sqrt{2\pi})^{-1}
\exp{\left[-(S^{m}_a-{\cal S})^2\sigma^{-2}/2\right]}$,
$\Delta s=2\times10^{-3}$ is the bin size), as predicted from Eq.
(\ref{Eq_slopes}).
%
%
%
The inset of Fig. \ref{Fig_3} shows $J^m_a$\cite{note-3} in units of
$J_0=-(C_R+C_G/2)/C_{\Sigma}$. Clearly,
this effect is negligible when compared with
$S^m_a$. Therefore, experimental measurements of
${\delta V^{th}}/{ \delta V_G}$ can accurately determine ${\cal
S}^{m}_a$ and hence the microscopic effects of the environmental fields
on the nanoisland energy levels; see Eq. (\ref{slopes}).
%
%
\section{Summary}
In summary, we have shown that coupling to environmental electric fields
gives rise to a unique fingerprint of the electron wavefunction at the
surface of a nanoscopic conductor that can be observed in tunneling
experiments.
While we illustrated this by considering Al/Al-oxide nanoislands
our findings rely only on the ability of ultrasmall {\em conducting} islands
to screen electric fields, and their large surface-to-volume ratio.
Therefore, other metallic nanoislands (noble and magnetic) and more
exotic nanoscopic conductors such as metallic carbon nanotubes
\cite{nanotubes} may also reveal the fingerprints of their
electron wavefunctions. We hope our results will motivate experimentalists to
look for these fingerprints.

\section*{Acknowledgments}

We thank John W. Wilkins for his helpful comments. This research
was funded by the Natural Science and Engineering Research Council of Canada (NSERC)
and the Canadian  Institute for Advanced Research (CIAR).

%
%
%
\begin{widetext}

%
\appendix
\section{\label{derivation_0}Derivation of Eq. (\ref{Eq_slopes})}
In Section \ref{environmental} we showed that the applied bias ($V$) and gate ($V_G$) voltages drive the
nanoisland to a potential $V_P$ that depends on the charge state ($Q_P$) of the nanoisland, 
and the values of $V$ and $V_G$. Applying {\em only} $V_G$ to a neutral nanoisland results in 
$V_P=(C_G/C_{\Sigma})V_G$. In this case ($V=Q_P=q=0$), the electrostatic potential energy ${\cal U}$ 
is given by:

%
\begin{equation}
\label{app:eq_11}
{\cal U}=-eV_G(C_G/C_{\Sigma})\left[-\alpha_L(\lambda d^L+2)^{-1}e^{-\lambda z^L_j}
-\alpha_R(\lambda d^R+2)^{-1}e^{-\lambda z^R_j}
+(C_{\Sigma}/C_G-1)\alpha_G(\lambda d^G+2)^{-1}e^{-\lambda z^L_j}\right]
\end{equation}
%

\noindent Note that Eq. (\ref{app:eq_11}) becomes {\em
different} for other $Q_P$, $V$, and $V_G$ values.

Within first order perturbation theory in ${\cal U}$, the microscopic contribution to the 
energy ${\cal E}^0_a(0,V_G)$ of level $|\psi_a\rangle$, $\varepsilon^0_a(0,V_G)$, is:

\begin{equation}
\label{app:eq_12}
{
\begin{array}{rl}
\varepsilon^0_a(0,V_G)= & \varepsilon^0_a(0,0)+\langle\psi^0_a|{\cal U}|\psi^0_a\rangle \\
 = & -eV_G(C_G/C_{\Sigma})\sum_j|\psi^0_a(j)|^2\left[-\alpha_L(\lambda d^L+2)^{-1}e^{-\lambda z^L_j}
-\alpha_R(\lambda d^R+2)^{-1}e^{-\lambda z^R_j}\right. \\
 & \left.+(C_{\Sigma}/C_G-1)\alpha_G(\lambda d^G+2)^{-1}e^{-\lambda z^L_j}\right] \\
\end{array}
}
\end{equation}

\noindent where $|\psi^0_a\rangle$ is an electronic eigenstate for $V_G=0$, as mentioned in the 
text (Sec. \ref{perturbation}), and the summation is performed over all atomic sites.

Equation (\ref{Eq_slopes}) in the text is straight forwardly
derived from Eq. (\ref{app:eq_12}).

%
%
%
\section{\label{derivation}Derivation of Eq. (\ref{slopes})}

As discussed in the text (Sec. \ref{perturbation}), $V^{th}_{L\rightarrow P}(V_G)$ is the 
minimum (threshold) value of the bias 
voltage---in the presence of an applied gate voltage $V_G$---at which an electron tunnels from 
electrode $L$ into an empty state $|\psi_a\rangle$ in the nanoisland ($P$).
This threshold value is given by:\cite{CB}

%
\begin{equation}
\label{app:eq_1}
{\cal E}_F =\varepsilon^{-}_a(V^{th}_{L\rightarrow P}(V_G),V_G)+ U_c -
[(C_R + C_G/2)/C_{\Sigma}]eV^{th}_{L\rightarrow P}(V_G)
-(C_G/C_{\Sigma})e V_G.
\end{equation}

\noindent After a simple manipulation, Eq. (\ref{app:eq_1}) leads to:

\begin{equation}
\label{app:eq_2}
\frac{\delta V^{th}_{L\rightarrow P}(V_G)}{ \delta V_G} =
-\frac{C_G}{C_R+C_G/2}
\left\{1+(-C_G/C_{\Sigma})^{-1}\left[\frac{
\varepsilon^{-}_a(V^{th}_{L\rightarrow P}(V_G+\delta V_G),V_G+\delta V_G)-
\varepsilon^{-}_a(V^{th}_{L\rightarrow P}(V_G),V_G)}{e\delta V_G}\right]
\right\}
\end{equation}
%

Then, we calculate within first order perturbation theory in ${\cal U}$ (setting $Q_P=-e$ and taking $V$
and $V_G$ as variables) the values 
of $\varepsilon^{-}_a(V,V_G)$. The latter leads to the following identity:

\begin{equation}
\label{app:eq_3}
\frac{
\varepsilon^{-}_a(V^{th}_{L\rightarrow P}(V_G+\delta V_G),V_G+\delta V_G)-
\varepsilon^{-}_a(V^{th}_{L\rightarrow P}(V_G),V_G)}{e\delta V_G}=\delta\varepsilon^{-}_a(V,V_G)/(e\delta V_G)=S^m_a+J^m_a\left(
\frac{\delta V^{th}_{L\rightarrow P}(V_G)}{ \delta V_G}\right)
\end{equation}  

\noindent where $J^m_a=\delta\varepsilon^{-}_a(V,V_G)/(e\delta V)$. By combining 
Eqs. (\ref{app:eq_2}) and (\ref{app:eq_3}) we get:

%
\begin{equation}
\label{app:eq_4}
\frac{\delta V^{th}_{L\rightarrow P}(V_G)}{\delta V_G} =
-\frac{C_G}{C_R+C_G/2}[1+S^m_a(-C_G/C_{\Sigma})^{-1}]
\{1+J^m_a[-(C_R+C_G/2)/C_{\Sigma})^{-1}]\}^{-1}
\end{equation}
%

\noindent Finally, by expanding the denominator of Eq. (\ref{app:eq_4})  in powers of 
$J^m_a[-(C_R+C_G/2)/C_{\Sigma}]$ we obtain Eq. (\ref{slopes}).

\end{widetext}

%


%
\begin{table*}[htb]
\begin{ruledtabular}
\begin{tabular}{cccccccccc|cc}
      label      & $D~(nm)$  & $h~(nm)$  & $N_{\Omega}/N_{{\cal V}}$ &
$E^0_{F}~(eV)$ &

$C_R~(aF)$ &
$C_L~(aF)$ & $C_G~(aF)$
      & $e/C_{\Sigma}~(mV)$
       & $e/2C_G~(mV)$ &   ${\cal S}$ &
$-C_G/C_{\Sigma}$
%
\\
\cline{1-12}
$A$ & 3.646   & 1.62  & 0.774 & 8.289 &    1.480 & 1.057 & 0.164 &
59.32
       & 488.47   & $-$0.0214  & $-0.0607$ \\
$B$ & 5.266   & 0.81  & 1.045 & 8.281 &    3.088 & 2.206 & 0.109 &
29.65  & 734.95  & $-0.0159$  & $-0.0202$
\end{tabular}
\end{ruledtabular}
\caption{\label{tab:Table_1}Transistor parameters: Diameter ($D$), height ($h$),
surface-to-volume ratio ($N_{\Omega}/N_{{\cal V}}$), Fermi energy of the
neutral isolated grain ($E^0_{F}$), capacitances, and charging and
degeneracy voltages.
${\cal S}$ is the average slope due to microscopic renormalization and
      $-C_G/C_{\Sigma}$ is the Orthodox model slope (see
text). }
\end{table*}


\begin{figure*}
\caption{(a) Schematic of the model transistors. The
island (black and white circles) is disc-shaped and is {\em
surrounded} by the gate ($G$). White circles indicate where environmental
electric fields are important.
(b) Sketch of the  deviation $\Delta V_z$
within the island. $O^{\prime}$
indicates its center.}
\label{Fig_1}
\end{figure*}

\begin{figure*}
\caption{Fingerprints of the electronic wave functions:
microscopic contribution to the energy levels of the transistors vs.
$V_G$ (at $V=0$) and $V$ (at $V_G=0$) with $Q_P=0$. Voltages are in
$mV$.}
\label{Fig_2}
\end{figure*}

\begin{figure*}
\caption{Histograms (broken lines) of
$S^{m}_a$ in {\em units} of the standard Orthodox theory slope: 
$-C_G/C_{\Sigma}$.
Gaussian distributions (smooth lines) corresponding to the calculated
${\cal S}$ and
$\sigma$ (see text). Inset: Histograms of $J^m_a$
in units of $J_0=-(C_R+C_G/2)/C_{\Sigma}$ for $A$ (bold line) and $B$ (thin
line).}
\label{Fig_3}
\end{figure*}


\begin{thebibliography}{40}


\bibitem{CB}
%
%
I. O. Kulik and R. I. Shekhter, Zh. Ekps. Teor. Fiz. {\bf 62}, 623
(1975)[Sov. Phys. JETP {\bf 41}, 308 (1975)]; K. K. Likharev, IBM J. Res.
Dev. {\bf 32}, 144 (1988); Proc. IEEE {\bf 87}, 606 (1999); D. V. Averin and
A. N. Korotkov, Zh. Eksp. Teor. Fiz. {\bf 97}, 1661 (1990) [Sov. Phys. JETP
{\bf 70}, 937 (1990)];
D. V. Averin and K. K. Likharev, in {\em Mesoscopic Phenomena in Solids}, Eds.
B. L. Altshuler, P. A. Lee, and R. A. Webb (Elsevier, Amsterdam, 1991); G.-L.
Ingold and Yu. Nazarov, in {\em Single Charge Tunneling}, Eds. H.
Grabert and M. H. Devoret,
(Plenum, NY, 1991);


\bibitem{set}
For reviews see: L. P. Kouwenhoven, D. G. Austing, and S. Tarucha,
Rep. Prog. Phys. {\bf 64}, 701 (2001);
Y. Takahashi, NTT Rev. {\bf 12}, 12 (2000); M. A. Kastner,
Rev. Mod. Phys. {\bf 64}, 849 (1992)

\bibitem{set1}
%
M. Ciorga, A. Wensauer, M. Pioro-Ladriere, M. Korkusinski, J. Kyriakidis, A.
S. Sachrajda, and P. Hawrylak, Phys. Rev. Lett. {\bf 88}, 256804 (2002);
%
J. Park, A. N. Pasupathy, J. I. Goldsmith,
C. Chang, Y. Yaish, J. R. Petta, M. Rinkoski,
J. P. Sethna, H. D. Abru\~na, P. L. McEuen, and D. C. Ralph,
Nature {\bf 417}, 722 (2002);
%
W. W. Liang, M. P. Shores, M. Bockrath, J. R. Long, H. Park,
Nature {\bf 417}, 725 (2002).
%
D. Goldhaber-Gordon, H. Shtrikman, D. Mahalu, D. Abusch-Magder, U. Meirav,
M. A. Kastner, Nature {\bf 391}, 156 (1998)

%
%

\bibitem{alhassid_RMP_2000} Y. Alhassid, Rev. Mod. Phys. {\bf 72}, 895 (2000);
A. D. Mirlin, Phys. Rep. {\bf 326}, 260 (2000)

\bibitem{RMT_reviews} K. B. Efetov, Adv. Phys. {\bf 32}, 53 (1983); W. P.
Halperin, Rev. Mod. Phys. {\bf 58}, 533 (1986)

\bibitem{brody_RMP_1981}  T. A. Brody,
J. Flores, J. B. French, P. A. Mello, A. Pandey, S. S. M. Wong,
Rev. Mod. Phys. {\bf 53}, 385 (1981)

\bibitem{narvaez_PRB_2002} G. A. Narvaez and G. Kirczenow, Phys. Rev. 
B {\bf 65},
121403(R) (2002); {\em ibid} {\bf 66}, 081404(R) (2002)


\bibitem{nanometals} D. C. Ralph, C. T. Black, and M. Tinkham,
Phys. Rev. Lett. {\bf 74}, 3241 (1995);
C. T. Black, D. C. Ralph, M. Tinkham,
Phys. Rev. Lett. {\bf 76}, 688 (1996);
D. Davidovi\'c and M. Tinkham, Phys. Rev. Lett. {\bf 83}, 1644 (1999);
Phys. Rev. B {\bf 61}, R16359 (2000);
S. Gu\'eron, M. M. Deshmukh, E. B. Myers, and D. C. Ralph,
Phys. Rev. Lett. {\bf 83}, 4148 (1999);
D. G. Salinas, S. Gu\'eron, D. C. Ralph, C. T.
Black, M. Tinkham, Phys. Rev. B {\bf 60}, 6137 (1999)

\bibitem{petta_PRL_2001} J. R. Petta, D.C. Ralph, Phys. Rev. Lett. 
{\bf 87}, 266801 (2001).

\bibitem{ralph_experiments}
M. M. Deshmukh, S. Kleff, S. Gu\'eron, E. Bonet, A. N. Pasupathy,
J. v. Delft, D. C. Ralph,
Phys. Rev. Lett. {\bf 87}, 226801 (2001);
M. M. Deshmukh, E. Bonet, A. N. Pasupathy, D. C. Ralph,
Phys. Rev. B{\bf 65}, 073301 (2002)

\bibitem{ralph_PRL_1997} D. C. Ralph, C. T. Black, and M. Tinkham, Phys. Rev.
Lett. {\bf 78}, 4087 (1997)


\bibitem{agam_PRL_1997} O. Agam, N. S. Wingreen, B. L. Altshuler, D. C.
Ralph, and M. Tinkham, Phys. Rev. Lett. {\bf 78}, 1956 (1997)

\bibitem{g-factor_models} P. W.
Brouwer, X. Waintal, and B. I. Halperin, Phys. Rev. Lett. {\bf 85},
369 (2000); K. A. Matveev, L. I. Glazman, and A. I. Larkin, Phys. Rev. Lett.
{\bf 85}, 2789 (2000)


\bibitem{theories} A. H. MacDonald, C. M. Canali, Solid State Comm. {\bf
119}, 253 (2001); C. M. Canali and A. H. MacDonald, Phys. Rev. Lett. {\bf 85},
5623  (2000); S. Kleff, J. v. Delft, M. M. Deshmukh, and D. C.
Ralph,  Phys. Rev. B 64, 220401(R) (2001);  I. L. Aleiner, P. W.
Brouwer and L. I. Glazman, Phys. Rep. {\bf 358}, 309 (2002);
E. Bonet, M. M. Deshmukh, and D. C. Ralph,
Phys. Rev. B{\bf 65}, 045317 (2002)

\bibitem{vondelft_PhysRep_2001} J. v. Delft and D. C. Ralph,
Phys. Rep. {\bf 345}, 61 (2001)

\bibitem{ashcroft_book} This assumption is justified within the
Thomas-Fermi screening model. For details of this topic see: {\em Solid State
Physics}, N. W. Ashcroft and N. D. Mermin (Saunders, 1976)

\bibitem{note_10} The derivation of this expression is straight forward.
It relies on (a) The assumption that the magnitude of the electric field
deep inside the electrodes ($L$, $R$, and $G$) and the nanoisland is negligibly
small and thus $V_L$, $V_R$, $V_G$, $V_P$ are equal to the values of the electrostatic
potential there;
and (b) The use of appropriate boundary  conditions for an electric field at the
metal/dielectric (the insulating oxide layer that separates the nanoisland from the
electrodes) interface. For details see: {\em Foundations of  Electromagnetic Theory},
J. R. Reitz, F. J. Milford,  and R. W. Christy (Adison-Wesley, 1979)

\bibitem{newns_PR_1970} D. M. Newns, Phys. Rev. B{\bf 1}, 3304 (1970)

\bibitem{note_14} Note that ${\cal U}$ depends on the charge state of the 
nanoisland ($Q_P$), applied bias ($V$) and gate ($V_G$) voltages.
  
\bibitem{note_2} The randomness of $|\psi^0_a(j)|^2$
       is also responsible
for the fluctuating g-factors of noble metal nanoislands
\cite{petta_PRL_2001,g-factor_models}.

\bibitem{note-3} In the linear regime $J^m_a=\delta
\varepsilon^{0}_a(V,0)/(e \delta V)$. Our simulations show that
this is a good approximation in general.

\bibitem{note_12} We briefly mention here details regarding the 
amount and type of disorder present in our model Al/Al-oxide nanoislands.
Further details may be found in Ref. \onlinecite{narvaez_PRB_2002}.
We first identify the atomic 
sites in the nanoisland that belong to the surface 
by looking at the number of nearest-neighbors that a given site has. In the cases where
this number is smaller than the coordination number in a face-centered-cubic lattice 
the atom is regarded as
a surface atomic site.  
In our model, the surface atoms represent the metal/oxide interface---the oxide coating. 
Hence, these atoms are {\em randomly} assigned to be either oxygen or
charged---due to oxidation---aluminum atoms. Here we assume for simplicity that equal
numbers of atoms of the two species are present at the surface.

\bibitem{note-2} These values of $d^i$ are physically reasonable choices.
Smaller values greatly {\em enhance} the microscopic effects.


\bibitem{averin_JETP_1990} D. V. Averin and A. N. Korotkov in
Ref. \onlinecite{CB}

\bibitem{nanotubes}
H. W. Ch. Postma, T. Teepen, Z. Yao, M. Grifoni, C. Dekker,
Science {\bf 293}, 76 (2001);
%
M. Bockrath, D. H. Cobden, P. L. McEuen, N. G. Chopra, A. Zettl,
A. Thess, R. E. Smalley, Science {\bf 275}, 1922 (1997)
%

\end{thebibliography}
\end{document}